\documentclass[aps,prb,twocolumn,superscriptaddress,
groupedaddress,10pt]{revtex4}
\usepackage{amsmath}
\usepackage{amssymb}
\usepackage{graphicx}
\usepackage{epsfig}
\usepackage{dcolumn}
\usepackage{bm}
\usepackage{blindtext}
\usepackage{natbib}
\usepackage{siunitx} 
\usepackage{multirow}
\usepackage{bbm}
\usepackage{subcaption} % 比 subfigure 更推荐
\usepackage{caption}
\usepackage[usenames,dvipsnames]{xcolor}
\usepackage{amsthm}
\usepackage{booktabs}
\usepackage{hyperref}
\usepackage{tikz}
\usetikzlibrary{calc}
\usepackage{mathtools}

\def\be{\begin{equation}} \def\ee{\end{equation}}
\def\bea{\begin{eqnarray}} \def\eea{\end{eqnarray}}

\def\nn{\nonumber}

\newcommand{\ket}[1]{| #1 \rangle}

\begin{document}
\title{Gapless and ordered phases in spin-1/2 Kitaev-XX-Gamma chain}
\author{Zebin Zhuang }
\affiliation{Department of Physics, Imperial College London, London SW7 2BZ, UK}
\author{Wang Yang }
\email{wyang@nankai.edu.cn}
\affiliation{School of Physics, Nankai University, Tianjin 300071, China}

\begin{abstract}

In this work, we study the spin-1/2 Kitaev chain with additional XX and symmetric off-diagonal Gamma interactions. 
By a combination of Jordan–Wigner transformation and density matrix renormalization group (DMRG) numerical simulations, we obtain the exact solution of the model and map out the phase diagram containing six distinct phases. 
The four gapped phases display  ferromagnetic and antiferromagnetic  magnetic orders along the $(1,1,0)$- and $(1,-1,0)$-spin directions,
whereas in the gapless phases, the low energy spectrum consists of two branches of helical Majorana fermions with unequal velocities. 
Transition lines separating different phases include deconfined quantum critical lines with dynamical critical exponent $z = 1$ and quadratic critical lines with $z = 2$. 
Our work reveals the rich interplay among symmetry, magnetic order, and quantum criticality  in the Kitaev–XX–Gamma chain.


%These results highlight the Kitaev–XX–Gamma chain as a minimal one-dimensional model capturing the interplay of symmetry, order, and criticality relevant to Kitaev materials.

\end{abstract}

%\pacs{75.10.Pq, 05.30.Rt, 71.10.Hf, 75.10.Jm}
\maketitle

%%%%%%%%%%%%%%%%%%%%%%%%%%%%%%%%%%%%%%%%%%%%%%%%%%%%%
\section{Introduction}

Quantum compass models \cite{Nussinov2015} are an important class of spin models where magnetic exchange interactions favor different spin components on different bonds,
sitting at the crossroads of spin-orbital physics and strongly correlated quantum magnetism. 
One-dimensional (1D) spin-1/2 quantum compass models are of special interests because of their often exact solvability via dualities and Jordan-Wigner transformations,
making possible an analytical understanding of various physical properties of the system, 
including excitation spectrum, correlation functions, entanglement entropies, to name a few \cite{Brzezicki2007,You2008,Eriksson2009,Sun2009,Mahdavifar2010,Motamedifar2011,Jafari2011,You2012,Liu2012_EPJB,Liu2012_PRB,You2014,Aziziha2013,Jafari2015,You2017,Laurell2023,Xu2025,Luo2025}.  
A typical quantum compass model in 1D is Kitaev's exactly solvable spin-1/2 model,
which has bond-dependent Ising interactions with two-site periodicity and an exponentially large ground state degeneracy \cite{Brzezicki2007}. 

Recently, there have been increasing research interests on the effects of an off-diagonal symmetric Gamma interaction $S_i^xS_j^y+S_i^yS_j^x$ \cite{Chaloupka2010}, which couples orthogonal spin directions, unlike the Ising type of interactions coupling the same directions of spins. 
The inclusion of the off-diagonal Gamma interaction is often indispensable in studying strongly correlated spin-orbit coupled magnets with edge-sharing octahedra, 
where the interplay between strong spin-orbit coupling and crystal fields lead to anisotropy in magnetic exchange interactions \cite{Nishimoto2015,Winter2016,Winter2017,Wang2017,Gohlke2018,Winter2019}. 
Therefore, the Gamma interaction is considered as a “real-world” correction to pure Kitaev/compass physics. 
In 1D, Ising and XY chains with an additional Gamma term have been considered,
in which rich physics including quantum phase transitions, quantum criticality, entanglements, and information scrambling have been investigated \cite{Kheiri2024,Zhao2022,Liu2021_PhysA,Liu2020_PRE,Mahdavifar2024_SciRep,Abbasi2025_SciRep,Abbasi2025_SciPostCore,Jin2025_PRE}.   
However, the 1D quantum compass model augmented with a Gamma term remains less explored. 
A prototypical  example of such 1D compass-Gamma model is the spin-1/2 Kitaev-XX-Gamma chain, 
which will be the focus of this work. 

It is worth to note that besides the above discussions,  the consideration  of 1D compass-Gamma model has its motivation rooted from a two-dimensional (2D) perspective as well, 
since in certain circumstances,  1D studies can provide hints for understanding 2D physics. 
The Kitaev spin-1/2 model  on the honeycomb lattice \cite{Kitaev2006} is an exactly solvable quantum compass model in 2D,
which hosts fractionalized anyonic excitations that can be used for realizing fault-tolerant topological quantum computations \cite{Kitaev2006,Nayak2008}. 
Real materials, however, inevitably host additional spin interactions allowed by lattice symmetries, which spoils the exact solvability of the  Kitaev honeycomb model and complicates the model both analytically and numerically \cite{Jackeli2009,Chaloupka2010,Winter2017,Trebst2022}.
In particular, the off-diagonal Gamma interaction has been shown to exist in real Kitaev materials with an interaction strength even comparable with the dominant Kitaev interaction,
and a typical model for describing 2D real Kitaev materials is the Kitaev-Heisenberg-Gamm model.

A productive route to a controlled understanding of the 2D case is to descend to 1D generalizations of the Kitaev model—obtained by restricting the honeycomb network to a chain and supplementing it with symmetry-allowed terms\cite{Yang2024a,Yang2020-1,Yang2022d,Yang2020b,Yang2021b,Luo2021,Luo2021b,Sorensen2021,Gordon2019,You2020b,Catuneanu2019,Agrapidis2019,Agrapidis2018,Sela2014,Yang2025,Yang2020,Yang2022a,Yang2022,Yang2022_2,Sorensen2023,Gruenewald2017}. 
 Despite their simplicity, they exhibit rich intrinsic physics, including nonsymmorphic symmetry group structures~\cite{Yang2022a}, nonlocal string orders\cite{Catuneanu2019,Luo2021}, solitonic excitations\cite{Sorensen2023}. Moreover, weakly coupled Kitaev chains can reconstruct hallmark 2D orders (e.g., zigzag and stripy magnetism~\cite{Yang2022_2,Yang2025}), thereby furnishing a quasi-1D bridge back to 2D Kitaev materials.
We note that the 1D Kitaev-XX-Gamma model considered in this work has a similar symmetry group as the Kitaev-Heisenberg-Gamma model.
The exact solvability of the spin-1/2 Kitaev-XX-Gamma chain makes a more thorough and comprehensive theoretical study be possible,
compared with the non-exactly-solvable Kitaev-Heisenberg-Gamma model.

In recent years, interest in one-dimensional quantum compass/Kitaev-type spin chains has been driven not only by their exact solvability and rich quantum critical behavior, but also by growing experimental motivation from real materials. In particular, the literature~\cite{Morris2025} demonstrated via THz spectroscopy that the quasi-1D magnet $\mathrm{CoNb_2O_6}$ is not simply described by the transverse-field Ising chain; instead, it realizes a bond-dependent “twisted Kitaev chain.” Their work shows that bond-anisotropic interactions induced by spin–orbit coupling and crystal symmetry can generate nontrivial domain-wall dynamics together, thereby directly linking 1D Kitaev/compass physics to $\mathrm{Co^{2+}}$-based materials. Inspired by this, we further consider a more realistic scenario relevant to edge-sharing octahedral magnets, and study a 1D Kitaev chain augmented by symmetry-allowed off-diagonal $\Gamma$ interactions and an additional XX exchange as “realistic corrections.”

In this work, we focus on the 1D spin-1/2 Kitaev-XX-Gamma model, 
which is an exactly solvable 1D quantum compass-Gamma model via Jordan-Wigner transformation. 
By combining Jordan-Wigner solution and DMRG numerical simulations, 
a variety of gapless and gapped phases  are revealed, occupying extended regions in the phase diagram. 
We find that the gapped phases have ferromagnetic (FM) and anti-ferromagnetic (AFM) orders along $(1,1,0)$- and $(1,-1,0)$-directions in the spin space, depending on the parameter regions;
while the gapless phases have a low energy spectrum consisting of two branches of helical Majorana fermions with distinct velocities. 
For the phase transitions,  the two AFM phases (similar for FM phases) are separated by a line of critical points with vanishing Gamma interaction, 
hence this critical line constitutes an example of deconfined quantum phase transition in 1D. 
On the other hand, the critical lines separating the gapped phases from the gapless regions have a quadratic low energy dispersion,
with dynamical critical exponent $z=2$. 

The rest of the paper is organized as follows. 
In Sec. \ref{sec:Ham}, the model Hamiltonian is introduced, and the associated symmetries are discussed. 
In Sec. \ref{sec:exact}, we apply the Jordan–Wigner transformation to exactly diagonalize the Hamiltonian and analyze the dispersion to delineate different phases. 
Sec. \ref{sec:phase_diagram} is devoted to a numerical study of the phase diagram, 
in which ground state degeneracies, correlation functions, central charges are calculated. 
Sec. \ref{sec:summary} briefly summarizes the main results of the paper.

\begin{figure}[h]
\begin{center}
\includegraphics[width=8.5cm]{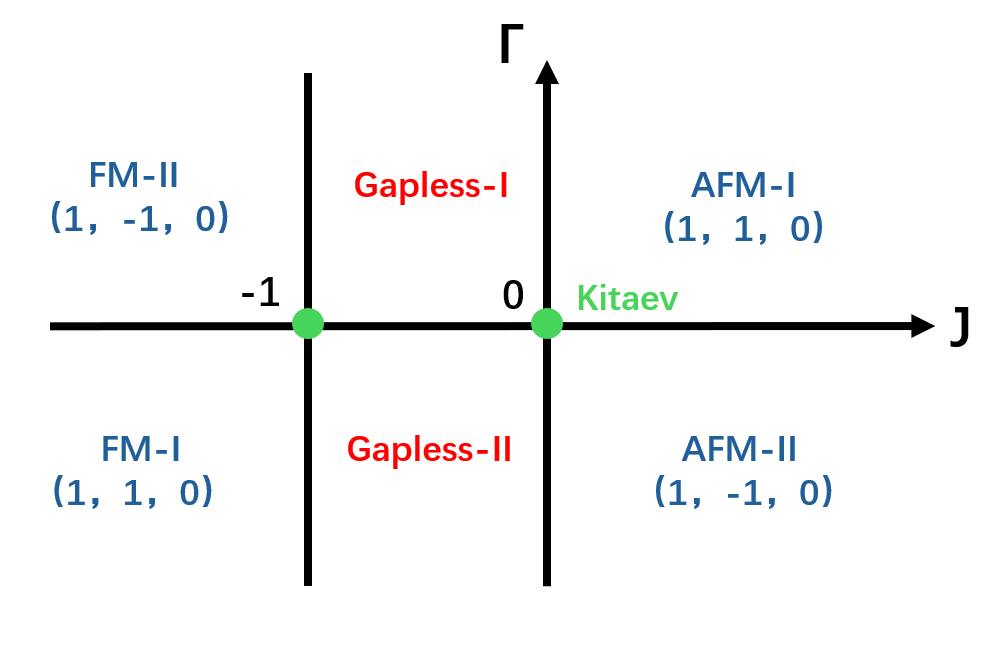}
   \captionsetup{justification=raggedright}
\caption{ Phase diagram of the 1D spin-1/2 Kitaev–XX–$\Gamma$ model with  $K=1$. AFM-I and FM-I denote antiferromagnetic and ferromagnetic phases polarized along the spin direction $(1,1,0)$, whereas AFM-II and FM-II denote the corresponding phases along $(1,-1,0)$. At the parameter points $(0,0)$ and $(-1,0)$ the model reduces to the pure Kitaev chain.
} 
\label{fig:phasediagram}
\end{center}
\end{figure}
\label{sec:intro}

%%%%%%%%%%%%%%%%%%%%%%%%%%%%%%%%%%%%%%%%%%%%%%%%%%%%%
\section{Model Hamiltonian and symmetries}
\label{sec:Ham}

%----------------------------------------------------------------------------------------------------------------
\subsection{Model Hamiltonian}
\label{subsec:Ham}

\begin{figure}[ht]
\begin{center}
\includegraphics[width=8cm]{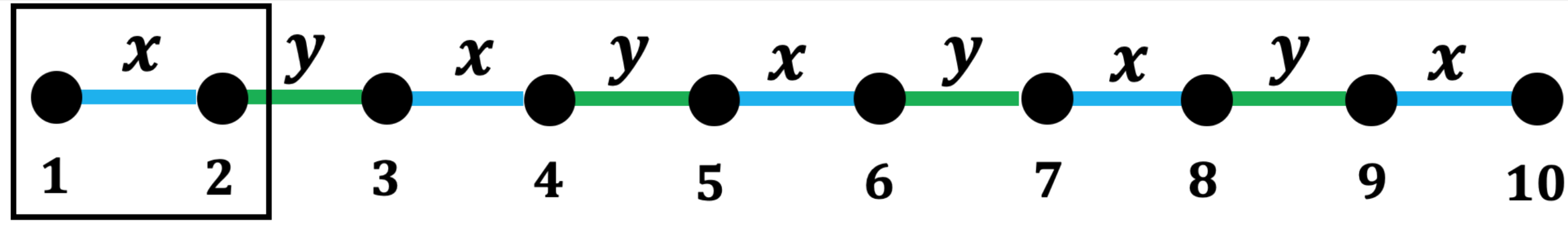}
   \captionsetup{justification=raggedright}
\caption{Bond pattern of the Kitaev-XX-Gamma chain.
} \label{fig:bonds}
\end{center}
\end{figure}

We consider the 1D spin-1/2 Kitaev-XX-Gamma model defined by the following Hamiltonian.
\begin{equation}
\begin{aligned}
  H=\sum_{<ij>\in\gamma\,\text{bond}}\big[ & KS_i^\gamma S_j^\gamma+ J(S_i^xS_j^x+S_i^yS^y_j)\\+&\Gamma(S^x_iS^y_j+S^y_iS^x_j) \big],
\label{eq:Ham}  
\end{aligned}
\end{equation}
in which $i,j$ are two sites of nearest neighbors;
$\gamma=x,y$ is the spin direction associated with the $\gamma$ bond connecting nearest neighboring  sites $i$ and $j$, with the pattern shown in Fig. \ref{fig:bonds}; $K$, $J$, and $\Gamma$ denote the Kitaev, XX, and the symmetric off-diagonal Gamma couplings, respectively.

%-------------------------------------------- 
%--------------------------------------------
%The choice of the coupling constant range in our model is demonstrated here.  
Since the special case $K=0$ has already been studied in the literature~\cite{Abbasi2025_SciRep,Abbasi2025_SciPostCore},  we restrict our analysis to the case $K \neq 0$. It is important to note that the two standard choices $K=\pm 1$ are related by a unitary two-sublattice rotation $U_2$, defined as:
\begin{eqnarray}
\text{Sublattice $1$}: & (x,y,z) & \rightarrow (-x^\prime,-y^\prime,z^{\prime}),\nn\\ 
\text{Sublattice $2$}: & (x,y,z) & \rightarrow (x^{\prime},y^{\prime},z^{\prime}),
\label{eq:4rotation}
\end{eqnarray}
where a spin rotation about the $\hat{z}$   by an angle $\pi$  is applied to every even site,
while the odd sites are leaving unchanged.  
The Hamiltonian then transforms as $H' = U_2 H U_2^{-1}$, which can be made explicit in the first unit cell containing sites $1$ and $2$ as follows:
\begin{flalign}
    H_{12}' &= -K S^x_1 S^x_2 -J(S^x_1 S^x_2+S^y_1 S^y_2)-\Gamma (S^x_1 S^y_2+ S^y_1 S^x_2),\nn\\
    H_{23}' &= -K S^y_2 S^y_3 -J(S^x_2 S^x_3+S^y_2 S^y_3)-\Gamma (S^x_2 S^y_3+ S^y_2 S^x_3).
\end{flalign}
Therefore, the $U_2$ rotation establishes an equivalence relation in parameter space,
\begin{equation}
  (K,J,\Gamma) \simeq (-K, -J, -\Gamma).
  \label{eq:equiv_K_pm}
\end{equation}
For a non-standard choice of $K$, we can map the parameters to the standard cases by redefining $\Gamma$ and $J$ in accordance with the substitutions  $\Gamma \rightarrow \Gamma/K$ and $J \rightarrow J/K$.  
Hence, because of the equivalence relation in Eq. (\ref{eq:equiv_K_pm}) and the rescaling of $\Gamma$ and $J$, without loss of generality, we can always fix $K=1$, and it suffices to study the parameter space spanned by $(J,\Gamma)$.

Consider the global spin rotation $R(\hat{x},\pi)$ applied to every site, 
where $R(\hat{n},\varphi)$ is used to denote a global spin rotation around the direction $\hat{n}$ by an angle $\varphi$.
Clearly, $R(\hat{x},\pi)$ flips $S_i^y$ to  $-S_i^y$ ($1\leq i\leq N$ where $N$ is the number of sites), 
and as a result, this operation only changes the sign of the $\Gamma$ coupling,
implying the following equivalence relation:
\begin{equation}
    (K,J,\Gamma)\;\simeq\;(K,J,-\Gamma).
    \label{er1}
\end{equation}

Next, we consider another two-sublattice rotation $U'_2$, defined as
\begin{eqnarray}
\text{Sublattice 1:} & (x,y,z) & \;\longrightarrow\; (-y^{\prime},\,x^{\prime},\,z^{\prime}), \nn\\ 
\text{Sublattice 2:} & (x,y,z) & \;\longrightarrow\; (y^{\prime},\,-x^{\prime},\,z^{\prime}),
\label{eq:2rotation}
\end{eqnarray}
which corresponds to applying the spin rotation around $z$-direction by an angle $\pi/2$ ($-\pi/2$) to every odd (even) site.
The transformed Hamiltonian in the first unit cell is then given by:
\begin{eqnarray}
    H_{12}' &=& KS^x_1 S^x_2 -(J+K)(S^x_1 S^x_2+S^y_1 S^y_2)\nn\\
    &&+\Gamma (S^x_1 S^y_2+ S^y_1 S^x_2),\nn\\
    H_{23}' &=& KS^y_2 S^y_3 -(J+K)(S^x_2 S^x_3+S^y_2 S^y_3)\nn\\
    &&+\Gamma (S^x_2 S^y_3+ S^y_2 S^x_3),
\end{eqnarray}
which reveals the following equivalence relation:
\begin{equation}
    (K,J,\Gamma)\simeq (K,-K-J, \Gamma). 
    \label{er2}
\end{equation}
%----------------------------------------------------------------------------------------------------------------
\subsection{Hamiltonian Symmetries}
The Hamiltonian in Eq. (\ref{eq:Ham}) is invariant under the following transformations:
\bea
T&:&(S_i^x,S_i^y,S_i^z)\rightarrow (-S_i^x,-S_i^y,-S_i^z)\nn\\
%T_{2a}&:&(S_i^x,S_i^y,S_i^z)\rightarrow (S_{i+2}^x,S_{i+2}^y,S_{i+2}^z)\nn\\
R(\hat{z},\pi)&:& (S_i^x,S_i^y,S_i^z)\rightarrow (-S_{i}^x,-S_{i}^y,S_{i}^z)\nn\\
R(\hat{n}_1,\pi)I&:&(S_i^x,S_i^y,S_i^z)\rightarrow (-S_{2-i}^y,-S_{2-i}^x,-S_{2-i}^z)\nn\\
%R(\hat{n}_2,\pi)I&:&(S_i^x,S_i^y,S_i^z)\rightarrow (S_{2-i}^y,S_{2-i}^x,-S_{2-i}^z)\nn\\
R(\hat{n}_1,\pi)T_a&:&(S_i^x,S_i^y,S_i^z)\rightarrow (-S_{i+1}^y,-S_{i+1}^x,-S_{i+1}^z),\nn\\
%R(\hat{n}_2,\pi)T_a&:&(S_i^x,S_i^y,S_i^z)\rightarrow (S_{i+1}^y,S_{i+1}^x,-S_{i+1}^z).
\label{eq:sym_ops}
\eea
in which $T$ is time reversal operation, $T_{na}$ is translation by $n$ lattice sites,   $I$ is spatial inversion with inversion center at site $1$, 
%$R(\hat{n},\theta)$ is a global spin rotation around $\hat{n}$ axis by angle $\theta$, 
and $\hat{n}_1$ is given by
%and $\hat{n}_2$ are given by
\bea
\hat{n}_1&=&\frac{1}{\sqrt{2}}(1,-1,0)^T.
%\hat{n}_2&=&\frac{1}{\sqrt{2}}(1,1,0)^T.
\eea
The symmetry group $G$ is generated by the operations in Eq. (\ref{eq:sym_ops}),
\begin{flalign}
G=\langle T, R(\hat{z},\pi), R(\hat{n}_1,\pi)I,R(\hat{n}_1,\pi)T_a\rangle.
\label{eq:sym_group}
\end{flalign}
Notice in particular that the operations $T_{2a},R(\hat{n}_2,\pi)I$ and $R(\hat{n}_2,\pi)T_a$ are also symmetries of the system where $\hat{n}_2=\frac{1}{\sqrt{2}}(1,1,0)^T$,
which can be expressed in terms of the operations in Eq. (\ref{eq:sym_ops}) as
$T_{2a}=[R(\hat{n}_1,\pi)T_a]^2$, $R(\hat{n}_2,\pi)I=R(\hat{z},\pi)\cdot R(\hat{n}_1,\pi)I$, $R(\hat{n}_2,\pi)T_a=R(\hat{z},\pi)\cdot R(\hat{n}_1,\pi)T_a$.

It is interesting to note the group structure of $G$.
Since $\langle T_{2a} \rangle$ is an abelian normal subgroup of $G$, it is legitimate to consider the quotient group $G/\langle T_{2a} \rangle$. 
Notice that in the sense of modulo $T_{2a}$, all the four generators of $G$ in Eq. (\ref{eq:sym_group}) square to the identity element and mutually commute,
therefore 
\bea
G/\langle T_{2a} \rangle\simeq (\mathbb{Z}_2)^4,
\eea
in which $\langle T, R(\hat{z},\pi), R(\hat{n}_1,\pi)I,R(\hat{n}_1,\pi)T_a\rangle/\langle T_{2a} \rangle \simeq(\mathbb{Z}_2)^4$. 
Namely, the symmetry group $G$ has a nonsymmorphic structure, characterized by the following short exact sequence
\bea
1\rightarrow \langle T_{2a} \rangle \rightarrow G\rightarrow (\mathbb{Z}_2)^4\rightarrow 1.
\eea

\section{Exact solution of the model}
\label{sec:exact}

\begin{figure*}[t]
  \centering

  % --- 第一行 (a)(b) ---
  \begin{subfigure}[t]{0.35\textwidth}
    \subcaption{}%
    \includegraphics[width=\linewidth]{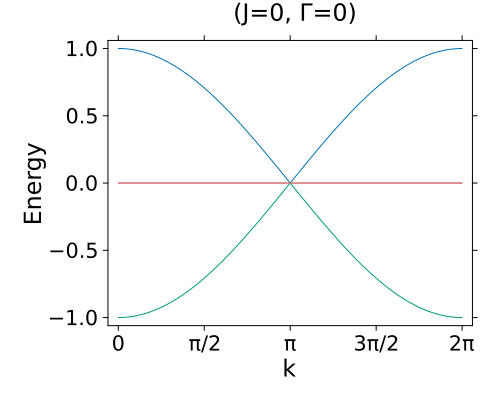}
    \label{fig:abcd-a}
  \end{subfigure}\hspace{0.01\textwidth}%
  \begin{subfigure}[t]{0.35\textwidth}
    \subcaption{}%
    \includegraphics[width=\linewidth]{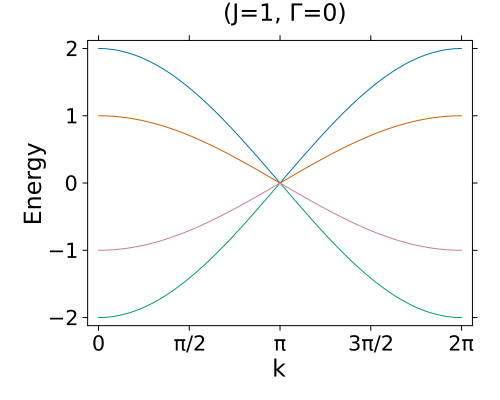}
    \label{fig:abcd-b}
  \end{subfigure}

  \vspace{-0.4em} % 上下两行更紧凑（-0.6~-1.0em 之间自己微调）

  % --- 第二行 (c)(d) ---
  \begin{subfigure}[t]{0.35\textwidth}
    \subcaption{}%
    \includegraphics[width=\linewidth]{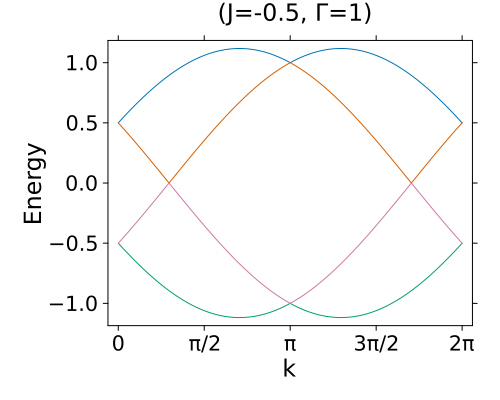}
    \label{fig:abcd-c}
  \end{subfigure}\hspace{0.01\textwidth}%
  \begin{subfigure}[t]{0.35\textwidth}
    \subcaption{}%
    \includegraphics[width=\linewidth]{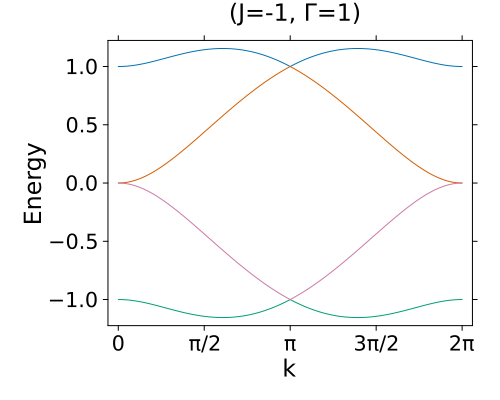}
    \label{fig:abcd-d}
  \end{subfigure}
   \captionsetup{justification=raggedright}
  \caption{Energy dispersions in the gapless phase for (a) $J=0,\Gamma=0$,
  (b) $J=1,\Gamma=0$, (c) $J=-0.5,\Gamma=1$, (d) $J=-1,\Gamma=1$.
  }
  \label{fig:abcd}
\end{figure*}

The Hamiltonian in Eq. (\ref{eq:Ham}) can be exactly solved via the Jordan–Wigner transformation. 
In this section, we discuss the exact solution of the model and the energy spectrum. 

\subsection{Jordan-Wigner transformation}

In Jordan-Wigner transformation, the spin-$\tfrac{1}{2}$ operators can be mapped into spinless fermionic operators as
\begin{equation}
    S^+_i= c^\dagger_i T_i,\quad
    S^-_i= c_i T_i, \quad
    S^z_i=c^\dagger_i c_i-\frac{1}{2},
\end{equation}
in which $c_i^\dagger$ creates a spinless fermion at site $i$,  $c_i^\dagger$ and $c_j$ obey the canonical anticommutation relation
\begin{equation}
    \{c^\dagger_i,c_j\}=\delta_{ij},
\end{equation}
and $T_i$ is the Jordan-Wigner string
\begin{equation}
    T_i=e^{i\pi \sum_{j=1}^{i-1}c^\dagger_jc_j} .
    \label{JWstring}
\end{equation}

Notice that $T_{N+1}$ is the fermion-parity operator, whose eigenvalues are $\pm 1$. 
We impose the periodic boundary condition $S^a_{N+1} = S^a_1$, where $N$ denotes 
the chain length (taken to be even throughout this work) and $a \in \{x,y,z\}$. 
Under the Jordan--Wigner transformation, the interaction at the boundary between sites $N$ 
and $1$ acquires an extra factor of $T_{N+1}$. 
Depending on the fermion parity, we impose the fermionic boundary condition
\begin{equation}
    c_{N+1} = \sigma\, c_1, \qquad \sigma = \pm 1.
    \label{bc}
\end{equation}
Consequently, a 
spin chain with periodic boundary conditions is mapped onto a fermionic chain whose 
boundary conditions are periodic (anti-periodic) in the even (odd) fermion-parity sector. 
We will only consider the periodic fermionic boundary condition ($\sigma=1$) in this section for simplicity. 
Calculations for the anti-periodic boundary condition can be carried out in a similar manner \cite{Wu2021}.

The Hamiltonian can be rewritten as the sum of contributions from odd and even bonds:
\begin{equation}
\begin{aligned}
H &= \sum_{i \in \text{odd}} \Big[ 
      K\,S_i^x S_{i+1}^x 
    + J \big( S_i^x S_{i+1}^x + S_i^y S_{i+1}^y \big) 
\\ &\quad + \Gamma \big( S_i^x S_{i+1}^y + S_i^y S_{i+1}^x \big) 
\Big] \\
&\quad + \sum_{i \in \text{even}} \Big[
      K\,S_i^y S_{i+1}^y 
    + J \big( S_i^x S_{i+1}^x + S_i^y S_{i+1}^y \big) 
\\ &\quad + \Gamma \big( S_i^x S_{i+1}^y + S_i^y S_{i+1}^x \big) 
\Big].
\end{aligned}
\end{equation}
Defining two sets of fermionic operators distinguished by the parity of the site index $i$,
\begin{equation}
    c_{2m} = b_m, \quad c_{2m-1}=a_m, \quad 1\leq m\leq N/2,
\end{equation}
we obtain the spinless fermion model after Jordan-Wigner transformation as
\begin{equation}
\begin{aligned}
     H= &\frac{1}{4}\sum_i \big[(K+2J)a_i^\dagger b_i+(K-2i\Gamma)a^\dagger_ib^\dagger_i\\&+ (K+2J)b^\dagger_i a_{i+1}-(K+2i\Gamma)b_i^\dagger a_{i+1}^\dagger+\text{H.c} \big].
\end{aligned}
\end{equation}

Performing the following Fourier transformation (where $a$ is the lattice constant)

\begin{equation}
\begin{aligned}
a_j &= \frac{1}{\sqrt{N/2}} \sum_{k} e^{-ik\cdot 2aj}\, a_k, \\
b_j &= \frac{1}{\sqrt{N/2}} \sum_{k} e^{-ik\cdot 2aj}\, b_k,
\end{aligned}
\end{equation}

the Hamiltonian can be expressed through $c_k,c_k^\dagger$ as
\begin{equation}
\begin{aligned}
H &= \sum_{k>0,\,k\in K_e } H_k \\
&= \sum_{k>0,\,k\in K_e} \big[ (A_k+B_k) a_k^\dagger b_k +(A_k-B_k) a_{-k}^\dagger b_{-k} \\&+(C_K+D_k)a^\dagger_{k}b^\dagger_{-k}+(C_k-D_k)a^\dagger_{-k}b^\dagger_{k}+\text{H.c} \big],
\end{aligned}
\label{eq:H_fourier}
\end{equation}
in which the coefficients $A_k,B_k,C_k,D_k$ are given by
\begin{flalign}
    A_k&= \frac{1}{4}(K+2J)(1+\cos(k)) ,\nn\\
    B_k&=\frac{i}{4}(K+2J)\sin(k),\nn\\
    C_k&=\frac{1}{4}\big[ (K-2i\Gamma)+(K+2i\Gamma)\cos(k)\big],\nn\\
    D_k&=\frac{1}{4}(iK-2\Gamma)\sin(k).
    \label{eq:ABCD}
\end{flalign}
In Eq. (\ref{eq:ABCD}), the lattice constant is set to $a=1/2$ for simplicity.

The fermionic boundary condition~\ref{bc} can be rewritten as 
\begin{equation}
    a_{N/2+1}=a_{1}.
\end{equation}
The above boundary condition restricts the allowed $k$-modes to a discrete set that depends on the parity of $N/2$.  
For the even case,
\begin{equation}
    K_{e}
    = \left\{\, \frac{2\pi n}{Na} \;\middle|\; n = \pm 1, \dots, \pm \frac{N}{4} \right\},
\end{equation}
and for the odd case,
\begin{equation}
    K_{e}
    = \left\{\, \frac{2\pi n}{Na} \;\middle|\; 
        n = 0, \pm 1, \dots, \pm \tfrac{1}{2}\!\left(\tfrac{N}{2}-1\right)
      \right\}.
\end{equation}
Here the subscript ``$e$'' in $K_{e}$ denotes the even-fermion-parity sector.

The Hamiltonian in Eq. (\ref{eq:H_fourier}) can be cast into the Bogoliubov-de
Gennes (BdG) form,
\begin{equation}
    H=\sum_{k>0,\,k\in K_e}\Psi_k^\dagger H_{BdG}(k)\Psi_k,
\end{equation}
where $\Psi_k$ is the Nambu spinor  defined as 
\begin{equation}
    \Psi_k=(a_k,b_k,a_{-k}^\dagger,b_{-k}^\dagger)^T,
\end{equation}
and the $4\times4$ Bogoliubov-de Gennes matrix $H_{\mathrm{BdG}}(k)$ is 
\begin{flalign}
&H_{BdG}(k)=\nn\\
&\left(\begin{array}{cccc}
0 & A_k{+}B_k & 0 & C_k{+}D_k\\
A^{*}_k{+}B_k^{*} & 0 & -C_k+D_k & 0\\
0 & -C^{*}_k+D_k^{*} & 0 & -A^{*}_k+B_k^{*}\\
C^{*}_k{+}D^{*}_k & 0 & -A_k+B_k & 0
\end{array}\right).
\label{BdG}
\end{flalign}

The BdG matrix satisfies particle–hole symmetry, meaning that there exists an anti-unitary operator $\mathcal{C}$ such that
\begin{equation}
    \mathcal{C} H_{\mathrm{BdG}}(k)\, \mathcal{C}^{-1} \;=\; - H_{\mathrm{BdG}}(-k).
    \label{phs}
\end{equation}
For the BdG matrix in Eq. (\ref{BdG}), the operator $\mathcal{C}$ is given by
\begin{equation}
    \mathcal{C} = P \,\mathcal{K},
\end{equation}
where $\mathcal{K}$ denotes complex conjugation,
and $P$ is a $4\times 4$ unitary matrix acting in the Nambu space given by
\begin{equation}
    P = 
    \begin{pmatrix}
    0 & 0 & 1 & 0 \\[4pt]
    0 & 0 & 0 & 1 \\[4pt]
    1 & 0 & 0 & 0 \\[4pt]
    0 & 1 & 0 & 0
    \end{pmatrix}.
\end{equation}
The particle--hole symmetry guarantees that the eigenvalues of the BdG Hamiltonian 
come in positive-negative pairs. For example, if $E$ is an eigenvalue at momentum $k$, then $-E$ is also an eigenvalue at momentum $-k$.

 The BdG Hamiltonian in Eq.(~\ref{BdG}) also admits spatial inversion symmetry, meaning that there exists a unitary operator $\mathcal{I}$ such that
\begin{equation}
\mathcal{I}\,H_{\mathrm{BdG}}(k)\,\mathcal{I}^{-1} = H_{\mathrm{BdG}}(-k).
\end{equation}
The corresponding $4\times 4$ representation is
\begin{equation} \mathcal{I} = \begin{pmatrix} 0 & 1 & 0 & 0 \\[4pt] 1 & 0 & 0 & 0 \\[4pt] 0 & 0 & 0 & 1 \\[4pt] 0 & 0 & 1 & 0 \end{pmatrix}. \end{equation}
This symmetry ensures that the energy spectrum satisfies $E_{l}(k)=E_{l}(-k)$, where $l$ denotes the band index.

By diagonalizing the BdG Hamiltonian~\eqref{BdG}, we obtain the energy spectra consisting of four branches 
$\{E_{+}(k), -E_{+}(-k), E_{-}(k), -E_{-}(-k)\}$, corresponding to the four eigenvalues of the matrix, 
where the dispersions are given by:
\bea
E_{\pm}(k)&=&|\sqrt{\Delta(k)}\,\pm \frac{1}{2}|K\cos(k/2)||\,,
        \label{dis}
\eea  
in which 
\bea
\Delta(k)&=&\Gamma^{2}\sin^{2}(k/2)   +\big(J+\frac{K}{2}\big)^2\cos^{2}(k/2).
\label{eq:Delta_k}
\eea
 We denote by 
$v_{lk}$ and $u_{lk}$ ($l=\pm$) the eigenvectors of $H_{\mathrm{BdG}}(k)$ 
corresponding to eigenvalues $E_l(k)$ and $-E_l(-k)$, respectively: 
\bea
H_{\mathrm{BdG}}(k)\,v_{lk} &=& E_l(k)\,v_{lk}, \\
H_{\mathrm{BdG}}(k)\,u_{lk} &=& -E_l(-k)\,u_{lk}.
\eea
Defining the operators through 
\bea
\gamma^\dagger_{lk}=\Psi_k^\dagger v_{lk},\qquad
\beta^\dagger_{lk}=\Psi_k^\dagger u_{lk},
\eea
From the paticle-hole symmetry, we can find that
\begin{equation}
    \beta_{lk}^\dagger=\gamma_{l,-k}
\end{equation}
Combining the $k$ and $-k$ sectors into a single sum over the momentum set $K_e$, the diagonalized Hamiltonian takes the compact form 
\begin{equation}
    H=\sum_{k\in K_e}\sum_{l=\pm}
    E_l(k)\left(\gamma_{lk}^\dagger\gamma_{lk}-\tfrac{1}{2}\right).
\end{equation}

The ground state $\ket{GS}$ is determined by the condition that it is annihilated by $\gamma_{lk}$, i.e.,
\bea
\gamma_{lk}\ket{GS}=0.
\eea

However, we note that the ground state obtained directly from the Jordan--Wigner transformation is 
non-degenerate and thus does not capture the full set of ground states of the original 
spin model. 
This discrepancy arises from taking only $\sigma=1$ in the fermionic boundary condition in Eq. (\ref{bc}), as considering only 
the periodic sector is insufficient; both periodic and anti-periodic sectors must be 
included to recover the correct ground state degeneracy.

\subsection{Gapless and gapped phases}
\label{subsec:exact_JW_phases}

To proceed, we fix $K=1$ as discussed in Sec. \ref{sec:Ham} and restrict our attention to the $(J,\Gamma)$ plane. 
Notice that $\Delta(k)$ in Eq. (\ref{eq:Delta_k}) is consistent with the equivalence relations in Eq. (\ref{er1}) and Eq. (\ref{er2}), and as a result, the dispersion in Eq. (\ref{dis}) is symmetric with respect to the $\Gamma=0$ and $J=-1/2$ lines in Fig. \ref{fig:phasediagram}.

\subsubsection{The gapless $\Gamma=0$ line}
\label{subsubsec:gapless_gamma_0_line}

Along the $\Gamma=0$ line, the two points $J=0$ and $J=-1$ are special, where the Hamiltonian reduces to the pure Kitaev model, and $E_-(k)$ vanishes identically, leading to the presence of Majorana flat bands as illustrated in Fig.~\ref{fig:abcd-a}.

When $\Gamma=0$ and away the aforementioned  two special points, 
the spin Hamiltonian reduces to the Kitaev-XX model without the $\Gamma$ interaction, 
and the energy dispersions in Eq. (\ref{dis}) always exhibit a zero point at wave vector $k_F=\pi$ as shown in Fig.~\ref{fig:abcd-b}.
Expanding the dispersions around \(k_F\) with \(q=k-k_F\) gives two helical branches of Majorana fermions with distinct Fermi velocities.
%\begin{equation}
%E_{a}(q)=\pm v_a\,|q| , \quad a=1,2.
%\label{eq:linearization}
%\end{equation}
In the low energy limit, namely, in the vicinity of $k_F=\pi$, both branches of the dispersion scale as
\begin{equation}
E(q)\sim q^{\,z},
\label{ex}
\end{equation}
where \(z=1\) is the dynamical critical exponent. 
Thus, along the \(\Gamma=0\) line the system is gapless and relativistic. 
  The two low-energy dispersions in Eq.~(\ref{ex}) define two Majorana fermions that exhibit an emergent Ising conformal symmetry, each associated with a corresponding set of Ising operator-product expansions (OPEs). Consequently, the low-energy theory consists of two branches of helical Majorana fermions whose combined central charge is $c=1$.

We note that as to be discussed in Sec. \ref{subsubsec:gapped_region} and Sec. \ref{subsec:ordered}, 
the regions $(J>0,\Gamma>0)$, $(J>0,\Gamma<0)$, $(J<-1,\Gamma>0)$ and $(J<-1,\Gamma<0)$ are gapped and have different magnetic orders. 
Therefore, the $(J>0,\Gamma=0)$ and $(J<-1,\Gamma=0)$ lines are continuous phase transition lines separating distinct ordered phases,
hence are lines of deconfined quantum critical points  beyond the conventional Landau paradigm of second order phase transitions \cite{Senthil2004a,Senthil2004,Sandvik2007,Shao2016}. 
%This implies that the K-XX-Gamma model has a Lieb-Schulz-Mattis-like behavior \textcolor{red}{(add citation)}, possibly protected by its nonsymmorphic symmetry group structure, which is worth for future investigations. 

\subsubsection{The gapless $-1<J<0$ region}
\label{sbsbsec:gapless}

When $\Gamma \neq 0$, the gapless region is given by $-1<J<0$. In this parameter range, the dispersion always exhibits two zero-energy points located at momentum $k_F$ and $2\pi-k_F$, where
\begin{equation}
    k_F = 2\cos^{-1}\!\left(\sqrt{\frac{\Gamma^2}{\Gamma^2 + |J(J+1)|}}\right) \, .
    \label{eq:kF}
\end{equation}
As shown in Fig.~\ref{fig:abcd-c}, in the gapless phase ($-1<J<0$), 
the low-energy dispersion is linear, giving rise to two helical 
branches of Majorana fermions with identical Fermi velocity but shifted in momentum. 
The dynamical critical exponent is $z=1$, with total central charge $c=1$.

\subsubsection{The gapless $J=0$ and $J=-1$ lines with dynamical critical exponent $z=2$}

On the phase transition lines at $J=-1$ and $J=0$ (except the two special points with $\Gamma=0$ mentioned before), however, 
the low energy spectrum has a quadratic dispersion, as shown in Fig.~\ref{fig:abcd-d}.
Hence the dynamical critical exponent is given by $z=2$. 
%, and the system does not have an emergent Lorentz symmetry at low energies. 
These two lines are phase boundaries separating the gapless $-1<J<0$ region from the gapped regions to be discussed in Sec. \ref{subsubsec:gapped_region}.

%This is known as the Lifshitz transition~\cite{You2020}, as illustrated in Fig.~\ref{fig:abcd-d}.

\subsubsection{The gapped $(J>0,\,\Gamma>0)$ and its unitarily  equivalent regions}
\label{subsubsec:gapped_region}

The $(J>0,\,\Gamma>0)$ has a gapped energy spectrum.
Because of the equivalence relations in Eq. (\ref{eq:equiv_K_pm}) and Eq. (\ref{er1}),
the regions $(J>0,\Gamma<0)$, $(J<-1,\Gamma>0)$, $(J<-1,\Gamma<0)$ are also gapped. 
Hence, it is enough to focus on the $(J>0,\,\Gamma>0)$ region. The energy gap in this region can be summarized compactly as
\begin{equation}
\begin{aligned}
   \Delta_g = &\min\left\{ J,\, \Gamma,\, \frac{\Gamma}{2}\sqrt{\tfrac{4D-1}{D}} \right\},  \\
   D =& \left(J+\tfrac{1}{2}\right)^2 - \Gamma^2.
\end{aligned}
\end{equation}

%For the other gapped region, the same relation for $\Delta_g$ continues to hold,  as it can be obtained through the corresponding symmetry operation of the model
\begin{figure*}[t]
  \centering

  % --- 第一行 (a)(b) ---
  \begin{subfigure}[t]{0.35\textwidth}
    \subcaption{}%
    \includegraphics[width=\linewidth]{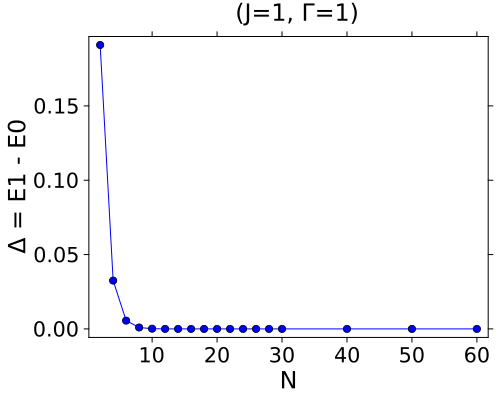}
    \label{fig:gsd-a}
  \end{subfigure}\hspace{0.01\textwidth}%
  \begin{subfigure}[t]{0.35\textwidth}
    \subcaption{}%
    \includegraphics[width=\linewidth]{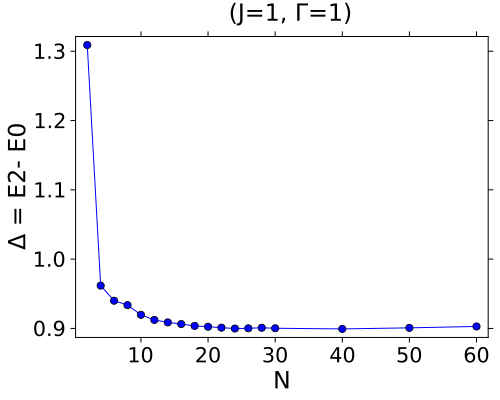}
    \label{fig:gsd-b}
  \end{subfigure}

\vspace{-0.4em}

  % --- 第一行 (a)(b) ---
  \begin{subfigure}[t]{0.35\textwidth}
    \subcaption{}%
    \includegraphics[width=\linewidth]{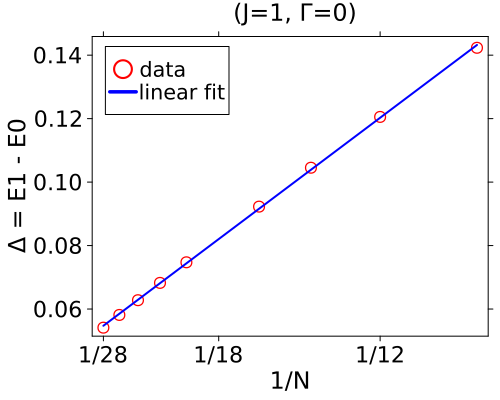}
    \label{fig:cp-a}
  \end{subfigure}\hspace{0.01\textwidth}%
  \begin{subfigure}[t]{0.35\textwidth}
    \subcaption{}%
    \includegraphics[width=\linewidth]{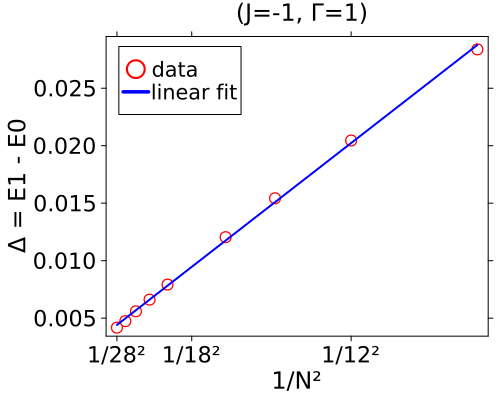}
    \label{fig:cp-b}
  \end{subfigure}
  
\captionsetup{justification=raggedright}
   \caption{
(a) $E_1-E_0$ vs. $N$ at $(J=1,\Gamma=1)$, (b) $E_2-E_0$ vs. $N$ at $(J=1,\Gamma=1)$,
(c) $E_1-E_0$ vs. $1/N$ at $(J=1,\Gamma=0)$, (d) $E_1-E_0$ vs. $1/N^2$ at $(J=-1,\Gamma=1)$,
where $E_0$, $E_1$, $E_2$ represent the lowest, next lowest, and third lowest energies, respectively. 
The chain length $N$ is varied from $2$ to $60$ in (a,b), and from $10$ to $28$ in (c,d).
The maximum bond dimension $m$ and truncation error $\epsilon$ in DMRG calculations are taken as $m= 1500$,  $\epsilon=10^{-12}$.  }
   \label{fig:gsd}
\end{figure*}
\subsubsection{Implication for phase diagram }

In summary, our analysis of dispersion relations and energy gaps imply that the $(J,\Gamma)$ plane is naturally partitioned into six distinct phases: four gapped phases and two gapless phases,
in accordance with Fig. \ref{fig:phasediagram}.
The vertical transition lines at $J=-1$ and $J=0$ separate the gapped and gapless regions, while the horizontal line at $\Gamma=0$ further divides the gapped region into four disconnected parts and the gapless region into two parts.

%-------------------------------------------- 

%--------------------------------------------

\section{Phase Diagram}
\label{sec:phase_diagram}

In the previous section, we identified six independent phases in the $(J,\Gamma)$ parameter space of the Kitaev-XX-$\Gamma$ model. In this section, we turn to an investigation of the physical properties within each of these phases. To this end, we employ the numerical DMRG simulations.

\subsection{Ordered Phase}
\label{subsec:ordered}

\subsubsection{Ground State Degeneracy}
\label{subsubsec:degeneracy}

%---------------------------------------------------------------------------------------------------------------
\begin{figure*}[t]
  \centering
  % --- 第一行 (a)(b) ---
  \begin{subfigure}[t]{0.35\textwidth}
    \subcaption{}%
    \includegraphics[width=\linewidth]{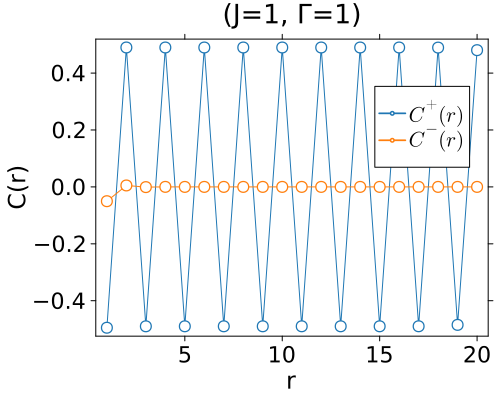}
    \label{fig:cf-a}
  \end{subfigure}\hspace{0.01\textwidth}%
  \begin{subfigure}[t]{0.35\textwidth}
    \subcaption{}%
    \includegraphics[width=\linewidth]{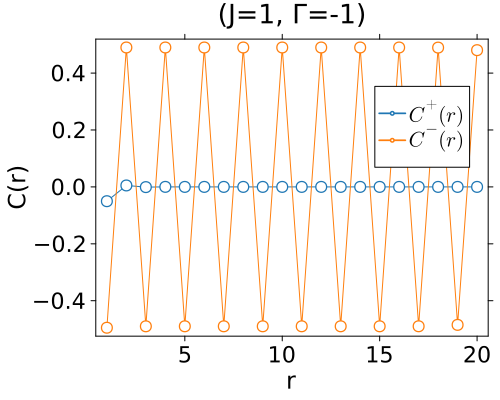}
    \label{fig:cf-b}
  \end{subfigure}

\vspace{-0.4em}

  % --- 第一行 (a)(b) ---
  \begin{subfigure}[t]{0.35\textwidth}
    \subcaption{}%
    \includegraphics[width=\linewidth]{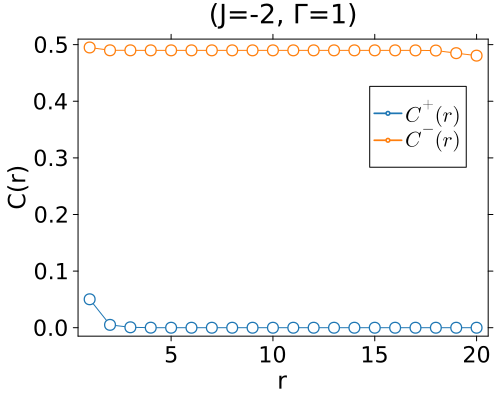}
    \label{fig:cf-c}
  \end{subfigure}\hspace{0.01\textwidth}%
  \begin{subfigure}[t]{0.35\textwidth}
    \subcaption{}%
    \includegraphics[width=\linewidth]{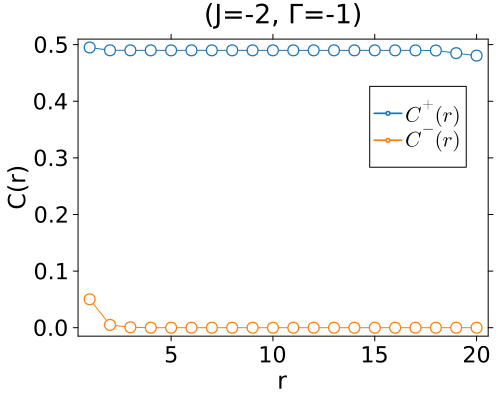}
    \label{fig:cf-d}
  \end{subfigure}
  
   \captionsetup{justification=raggedright}
   
   \caption{
   Correlation functions $C^{\pm}(r)$ as a function of $r$ for (a) $(J=1,\Gamma=1)$, (b) $(J=1,\Gamma=-1)$, (c) $(J=-2,\Gamma=1)$, and (d) $(J=-2,\Gamma=-1)$.
   DMRG numerics are performed on systems of $N=100$ sites using open boundary conditions.
   The maximum bond dimension $m$ and truncation error $\epsilon$ in DMRG calculations are taken as $m= 1500$,  $\epsilon=10^{-12}$.}
   \label{fig:cf}
\end{figure*}
%---------------------------------------------------------------------------------------------------------------
As mentioned earlier, the ground state in the periodic sector of the Jordan-Wigner fermion does not capture all possible features of the original spin model. 
In particular, if the ground states of the system are degenerate, then spontaneous symmetry breaking can occur, 
implying the existence of local order parameters. 
Therefore, the next step is to determine whether the ground state is degenerate.

In a finite size system, there is an exponentially small energy splitting $\Delta(N)$ between the degenerate ground states, having the following scaling form
\begin{equation}
\Delta(N) = A e^{-N/\xi},
\end{equation}
where $N$ is the system size, $A$ is a constant, and $\xi$ denotes the correlation length. 
In this case, the finite size energy gap decreases exponentially with increasing chain length and eventually closes in the thermodynamic limit($N\to\infty$).

By contrast, the excitation gap, defined as the energy difference between the first excited state and the ground states, behaves as
\begin{equation}
\Delta_{e}(N) = \Delta_{\infty} + B\,e^{-N/\xi},
\end{equation}
where $\Delta_{\infty}$ represents the excitation gap in the thermodynamic limit, and $B$ is a constant. Here, the gap converges to a finite constant as the system size increases. Therefore, by observing the scaling behavior of the energy gap in DMRG simulations, we can distinguish between degenerate and non-degenerate ground states: if the gap tends to close exponentially rapidly with increasing $N$, the ground state is degenerate.

The simulation results in Fig.~\ref{fig:gsd} are consistent with degenerate ground states. 
In the gapped phase, the finite-size energy gap exhibits exponential decay with system size. Specifically, Fig.~\ref{fig:gsd-a} shows the gap between the first ground state and the lowest-lying orthogonal state with respect to it. This splitting decreases rapidly and vanishes in the thermodynamic limit within numerical precision ($\sim5\times 10^{-14}$), demonstrating that the two states form a degenerate ground-state manifold. By contrast, Fig.~\ref{fig:gsd-b} displays the gap between the first ground state and the second lowest-lying orthogonal state. This quantity converges to a finite value $\Delta_{\infty}\approx0.9$, confirming that it corresponds to the true excitation gap of the system. Therefore, we conclude that the ground state in the $(J>0,\Gamma>0)$ phase is two-fold degenerate. We have also investigated other gapped phases, and the analysis indicates that their ground states are likewise two-fold degenerate,
which is as expected because of the equivalence relations in Eqs. (\ref{er1},\ref{er2}).

On the other hand, along the phase transition lines,
the ground state is expected to be non-degenerate,
and the excitation gap follows a power-law scaling,
\begin{equation}
\Delta(N) \sim N^{-z},
\end{equation}
where the critical exponent $z$ is defined as in Eq.~\ref{ex}. As shown in Fig.~\ref{fig:cp-a} and Fig.~\ref{fig:cp-b}, we examine two representative points on the transition line $\Gamma=0$ ($z=1$) and $J=-1$ ($z=2$). Plotting the energy gap against $1/N^{z}$ yields a clear linear behavior, confirming the expected scaling form.

\subsubsection{Correlation Functions}

We will show that the system exhibit FM or AFM order along $(1,1,0)$- or $(1,-1,0)$-directions, depending on the parameter region.
In this subsection, correlation functions will be investigated. 

Define operators $\Lambda_i^{\pm}$ at site $i$ as
\bea
\Lambda^{\pm}_i = S^x_i \pm S^y_i.
\eea
We examine the corresponding two-point correlation function,
\begin{equation}
    C^\lambda(r)=\langle \Lambda^{\lambda}_i \Lambda^{\lambda}_{i+r}\rangle,
    \label{correlation}
\end{equation}
where $\lambda=\pm$, and $\langle...\rangle$ denotes the expectation value with respect to the ground state. 
If $C^\lambda(r)$ saturates to a finite value as $r\rightarrow\infty$, 
the system has a magnetic order characterized by $\Lambda^{\lambda}_i$. 

The results of DMRG simulations for $C^\lambda(r)$ are shown in Fig.~\ref{fig:cf}. 
In the \(J>0\) region, as illustrated in Fig.~\ref{fig:cf-a} (Fig.~\ref{fig:cf-b}), 
the correlation function \(C^+(r)\) ($C^-(r)$) for $\Gamma>0$ ($\Gamma<0$) exhibits a staggered structure as a function of \(r\),
whereas \(C^-(r)\) ($C^+(r)$) vanishes when $r$ becomes large. 
These behaviors indicate that the system develops antiferromagnetic (AFM) order along the \((1,1,0)\)- and \((1,-1,0)\)-directions,
in the $(J>0,\Gamma>0)$ and $(J>0,\Gamma<0)$ regions, respectively.
These two phases are denoted as AFM-I and AFM-II in the phase diagram in Fig. \ref{fig:phasediagram}.
The corresponding order parameters $\mathcal{O}^{s,\pm}_i$ are given by
\begin{equation}
    \mathcal{O}^{s,\pm}_i = (-1)^i \big(S^x_i \pm S^y_i\big), 
    \label{o1}
\end{equation}
in which ``$+$" for AFM-I phase and ``$-$" for AFM-II phase,
and the superscript \(s\) in $\mathcal{O}^{s,\pm}_{i}$ is used to denote the staggered nature of the operator. 
Moreover, the equivalence relation Eq. (\ref{er1}) implies that these two order parameters must be transformed into each other via the global spin rotation $R(\hat{x},\pi)$ as
\begin{equation}
U\mathcal{O}^{s,+}_iU^{\dagger} =\mathcal{O}^{s,-}_i,
\end{equation}
where $U$ denotes the unitary operator corresponding to $R(\hat{x},\pi)$, the above transformation can be easily verified. 

In the region \(J<-1\), as illustrated in Fig.~\ref{fig:cf-c} and Fig.~\ref{fig:cf-d}, 
the correlation functions $C^-(r)$ ($C^+(r)$) converges to a uniform nonzero value as \(r \to \infty\) for $\Gamma>0$ ($\Gamma<0$). 
These two phases are denoted as FM-I and FM-II in the phase diagram in Fig. \ref{fig:phasediagram}.
The corresponding order parameters are
\begin{equation}
    \mathcal{O}^{\pm}_i = S^x_i \pm S^y_i, 
    \label{o2}
\end{equation}
in which ``$+$" for $\Gamma<0$ and ``$-$" for $\Gamma>0$.
Furthermore, the equivalence relation in Eq. (\ref{er2}) implies that the AFM and FM order parameters can be transformed into each other via a two-sublattice rotation \(U_2'\) defined in Eq.~\ref{eq:2rotation},  as
\begin{equation}
U_2'\mathcal{O}^{s,\pm}_iU_2^{\prime\dagger} =\mathcal{O}^{\mp}_i,
\end{equation}
which can be straightforwardly verified.

It is worth to note that we evaluate the spin correlation functions numerically using DMRG, however, it is possible to determine it analytically via the Jordan-Wigner transformation directly. Under the Jordan–Wigner transformation, spin operators acquire non-local string factors as given in Eq.(\ref{JWstring}). For transverse components of the spin, these strings do not cancel completely and leave behind a long product of Majorana fermion operators. Because the fermionic Hamiltonian is quadratic, the expectation value of this long Majorana chain can be reduced, through Wick’s theorem, to the Pfaffian of an antisymmetric matrix constructed from two-point fermionic correlators. In a translation-invariant system, the matrix elements depend only on the relative separation of the sites involved, so the resulting Pfaffian is equivalent to the determinant of a Toeplitz or block-Toeplitz matrix. In the thermodynamic limit, these become infinite-dimensional Toeplitz determinants whose asymptotic behaviour governs the spin correlation functions of the large-distance~\cite{Montroll1963,Lieb1961,Barouch1971}. The resulting analytical solution is expected to agree with our DMRG calculations.

\subsubsection{Symmetry breaking patterns}

We now analyze the symmetry-breaking patterns in AFM-I and AFM-II phases. 
Analysis of symmetry breaking patterns  can be carried  out in a similar manner in the FM-I and FM-II phases. 

The symmetry generators of the Hamiltonian are given in Eq.~(\ref{eq:sym_ops}). 
It can be verified that the unbroken symmetry groups $G_{\mathrm{AFM\text{-}I}}$ and $G_{\mathrm{AFM\text{-}II}}$ of the magnetic orders in the AFM-I and AFM-II phases are given by 
\begin{equation}
    \begin{aligned}
        G_{\mathrm{AFM\text{-}I}}  &= \langle \, T \cdot R(\hat{z},\pi),\; R(\hat{n}_2,\pi)T_a,\; R(\hat{n}_1,\pi)I \, \rangle,\\
        G_{\mathrm{AFM\text{-}II}} &= \langle \, T \cdot R(\hat{z},\pi),\; R(\hat{n}_2,\pi)I,\; R(\hat{n}_1,\pi)T_a \, \rangle .
    \end{aligned}
    \label{sym_gsg}
\end{equation}
Using the group relations, one finds that the two of the nonsymmorphic generators in Eq. (\ref{sym_gsg}) square to a 
two-site translation,\
\begin{equation}
    [R(\hat{n}_2,\pi)T_a]^2 = T_{2a}, \qquad
    [R(\hat{n}_1,\pi)T_a]^2 = T_{2a},
\end{equation}
while the remaining generators square to the identity. Thus, the normal subgroup $\langle T_{2a}\rangle$ is contained in both $G_{\mathrm{AFM\text{-}I}}$ and $G_{\mathrm{AFM\text{-}II}}$. In the sense of modulo $T_{2a}$, all the three generators of $G_{AMF-I}$ and $G_{AMF-II}$ in Eq. (\ref{sym_gsg}) square to the identity element and mutually commute, 
yielding the following quotient structure
\begin{equation}
    \begin{aligned}
        G_{\mathrm{AFM\text{-}I}}/\langle T_{2a}\rangle  &\cong (\mathbb{Z}_2)^3,\\
        G_{\mathrm{AFM\text{-}II}}/\langle T_{2a}\rangle &\cong (\mathbb{Z}_2)^3.
    \end{aligned}
\end{equation}

Since the Hamiltonian symmetry group satisfies 
$G/\langle T_{2a}\rangle \cong (\mathbb{Z}_2)^4$, the ground-state degeneracy in both AFM-I and AFM-II phases are given by
\begin{equation}
    |(\mathbb{Z}_2)^4|/|(\mathbb{Z}_2)^3|
    =2,
\end{equation}
which agree with the numerical results in DMRG simulation discussed in Sec. \ref{subsubsec:degeneracy}. 
Thus, both AFM-I and AFM-II phases exhibit the following symmetry-breaking patterns,
\begin{equation}
    (\mathbb{Z}_2)^4 \rightarrow (\mathbb{Z}_2)^3 .
\end{equation}
However, although symmetry breaking in both AFM-I and AFM-II phases exhibit the structure $\mathbb{Z}_2^4 \rightarrow \mathbb{Z}_2^3$, they are different symmetry breaking patterns as the unbroken symmetry group $\mathbb{Z}_2^3$ are different for the AFM-I and AFM-II phases as can be seen from Eq. (\ref{sym_gsg}), thereby indicating that these two phases are distinct phases in the framework of Landau symmetry breaking theory.

\subsubsection{Deconfined quantum critical line}

\begin{figure}[h]
\centering
\includegraphics[width=8.5cm]{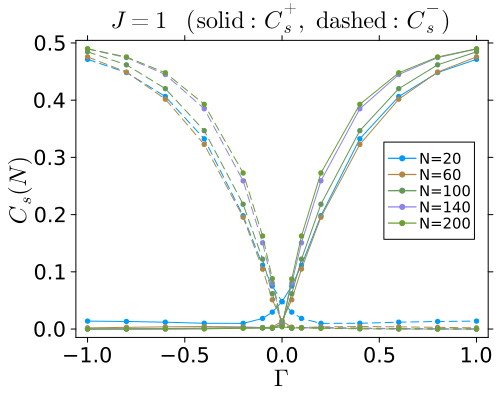}
 \captionsetup{justification=raggedright}
\caption{The spin correlation function defined in Eq.~\ref{eq:cf_2} is evaluated for several chain lengths, as shown in the figure. The coupling is fixed at $J = 1$, while $\Gamma$ is varied from $1$ to $-1$ in 11 uniform steps. The correlation distance is chosen as $r = N$. DMRG simulations are performed using open boundary conditions. The maximum bond dimension $m$ and truncation error $\epsilon$ in DMRG calculations are taken as $m= 1500$,  $\epsilon=10^{-12}$.} 
\label{fig:finitesize}
\end{figure}

To further numerically confirm the existence of the line of deconfined quantum critical points discussed in Sec. \ref{subsubsec:gapless_gamma_0_line}, 
we examine the following long-distance correlation functions associated with the AFM-I and AFM-II order parameters using DMRG numerics,
\begin{equation}
    C^{\pm}_s(N)= \langle \mathcal{O}^{s,\pm}_{1}\mathcal{O}^{s,\pm}_{N}\rangle.
    \label{eq:cf_2}
  \end{equation}
Fixing the coupling J and varying $\Gamma$, we monitor the evolution of this correlation function. In the presence of a continuous quantum phase transition, the order parameter is expected to shrink continuously to zero without exhibiting discontinuous jumps when approaching the critical point. 
Our numerical simulations, summarized in Fig.~\ref{fig:finitesize}, are consistent with this expectation. 
The finite-size effect, as shown in the Fig.~\ref{fig:finitesize}, indicates that the spin correlation function becomes increasingly size-independent as N grows, and the curves for different system sizes gradually collapse onto a single profile. 
This behavior rules out any discontinuous behavior of the order paramerters typical of a first-order transition and strongly suggests that the AFM-I to AFM-II transition is continuous, supporting the interpretation of the phase boundary as a candidate deconfined quantum critical line. A similar argument also applies to the transition between the FM-I and FM-II phases.

\subsubsection{Energy-Field Relation}

In order to further confirm the nature of the ordered phases, we now examine the energy-field relations. 
It is sufficient to focus on the FM-I phase, since the other phases are related to it by $U$ and $U_2^\prime$ discussed previously. 
Specifically, we consider the response of the system to the following types of external fields:
\begin{equation}
\begin{gathered}
-\frac{1}{\sqrt{2}}\,h_{x+y}\sum_i \big(S_i^x+S_i^y\big), \\
-\frac{1}{\sqrt{2}}\,h_{x-y}\sum_i \big(S_i^x-S_i^y\big), \\
-h_x\sum_i S^x_i ,\\
-h_y\sum_i S^y_i ,
\end{gathered}
\label{sf}
\end{equation}
which are added individually to the Hamiltonian in Eq. (\ref{eq:Ham}). 
If the system is ordered, then arbitrarily small field is able to induce a nonzero expectation value of the order parameter in the thermodynamic limit, and the relation between ground state energy change and the applied field will be linear in the limit of small fields. 
We monitor the ground-state energy as a function of the applied fields and identify which field induces the strongest response.

\begin{figure}[h]
\centering
\includegraphics[width=7.5cm]{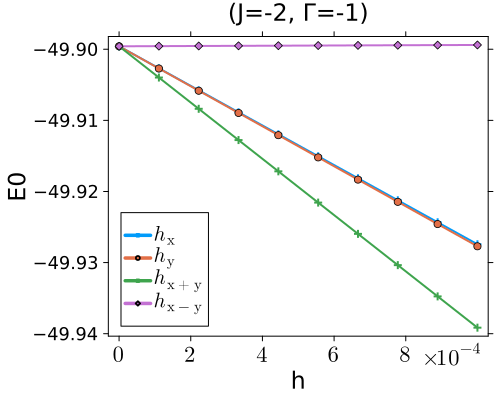}
 \captionsetup{justification=raggedright}
\caption{
Ground state energy as a function of the applied fields,
where the strength of the fields is varied from $0$ to $10^{-3}$ in 10 steps. 
DMRG simulations are performed in the FM-I phase with system size $N=80$ under open boundary conditions. 
 The maximum bond dimension $m$ and truncation error $\epsilon$ in DMRG calculations are taken as $m= 1500$,  $\epsilon=10^{-12}$.
} 
\label{fig:KHG_EnergyField}
\end{figure}

DMRG numerical results for all four types of fields in Eq. (\ref{sf}) are shown in Fig.~\ref{fig:KHG_EnergyField},
in which the responses to $h_x$ and $h_y$ collapse to a common line and cannot be clearly distinguished. 
The ground state energy responds linearly to all applied fields except $h_{x-y}$ (having no response), as expected in the small-field regime. 
Among the different perturbations, the field applied along the \((1,1,0)\)-direction, namely \(h_{x+y}\), produces a larger response compared to the fields applied along the \((1,0,0)\)- and \((0,1,0)\)-directions for \(h_x\) and \(h_y\). 
This confirms that system has an FM order along $(1,1,0)$-direction in the parameter region $(J<-2,\Gamma<0)$.

\begin{figure*}[t]
  \centering
  % --- 第一行 (a)(b) ---
  \begin{subfigure}[t]{0.4\textwidth}
    \subcaption{}%
    \includegraphics[width=\linewidth]{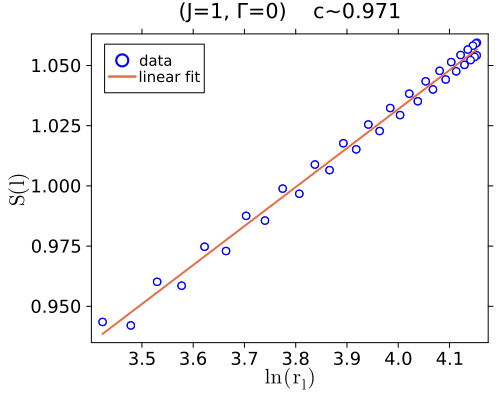}
    \label{fig:cc-a}
  \end{subfigure}\hspace{0.01\textwidth}%
  \begin{subfigure}[t]{0.4\textwidth}
    \subcaption{}%
    \includegraphics[width=\linewidth]{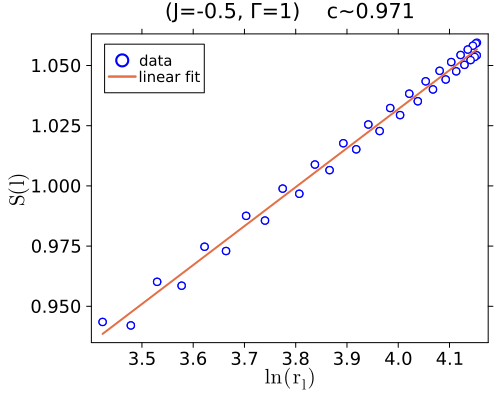}
    \label{fig:cc-b}
  \end{subfigure}
   \captionsetup{justification=raggedright}
   \caption{
   Extraction of central charge from a linear fit of entanglement entropy $S(l)$ vs. $\ln(r_l)$ at (a) \((J=1,\,\Gamma=0)\) and (b) \((J=-0.5,\,\Gamma=1)\),
   where $r_l = \frac{ N}{\lambda\pi} \sin \left(\frac{\pi l}{N}\right)$. 
DMRG simulations are performed on an open chain with \(N=100\) sites, where the outermost \(15\%\) of the data points near the edges are discarded. 
The maximum bond dimension and truncation error in the calculations are set to \(m=1500\) and \(\epsilon=10^{-12}\), respectively.}
   \label{fig:cc}
\end{figure*}

\subsection{Gapless Phase}

As discussed in Sec. \ref{subsec:exact_JW_phases},
the system is gapless within the range $-1\leq J\leq 0$. 
For brevity, we refer to the two gapless phases as Gapless-I for $\Gamma>0$ and Gapless-II for $\Gamma<0$,
as shown in Fig. \ref{fig:phasediagram}.

\subsubsection{Central Charge}

When the low-energy physics of the system is governed by a critical theory, 
the entanglement entropy \(S(l)\) of a subregion of length \(l\) in a finite chain of total length \(N\) is predicted by conformal field theory (CFT) to scale as \cite{Calabrese2009,Ejima2011}
\begin{align}
    S(l) = \lambda \frac{c}{3} \ln (r_l) + \cdots, 
    \label{eq:SL_formula}
\end{align}
in which \(c\) is the central charge, 
\(\lambda=1\) (\(\lambda=1/2\)) corresponds to periodic (open) boundary conditions \cite{Laflorencie2006}, 
$r_l = \frac{ N}{\lambda\pi} \sin \left(\frac{\pi l}{N}\right)$ which reduces to $l/\lambda$ in the limit $l\ll N$,
and ``\(\cdots\)" denotes subleading corrections. 
As mentioned in Sec.~2, the phase transition line \(\Gamma=0\) in Fig. \ref{fig:phasediagram} (except at two special points $J=0$ and $J=-1$) realizes a critical theory described by a \(c=1\) CFT, as do the two gapless phases.

Fig.~\ref{fig:cc} shows the DMRG numerical results for $S(l)$ as a function of $\ln r_l$ at two representative points \((J=1,\,\Gamma=0)\) on the transition line and \((J=-0.5,\,\Gamma=1)\) in the Gapless-II phase. 
Under open boundary conditions, the entanglement entropy \(S(l)\) exhibits alternating oscillations, 
which can be clearly seen in both panels of Fig.~\ref{fig:cc}. 
%The simulation data split into two distinct linear branches, reflecting the even-odd dependence of \(l\). 
To fit the central charge more precisely, we discard data points near the edges so that boundary effects can be reduced.
The fitted slopes in Fig.~\ref{fig:cc} yield central charge estimates of approximately \(c\simeq0.971\) at both representative points. 
With increasing system size \(N\), we expect the fitted central charge to converge to the theoretical value \(c=1\).

\subsubsection{Correlation functions}

\begin{figure}[h]
\centering
\includegraphics[width=8.5cm]{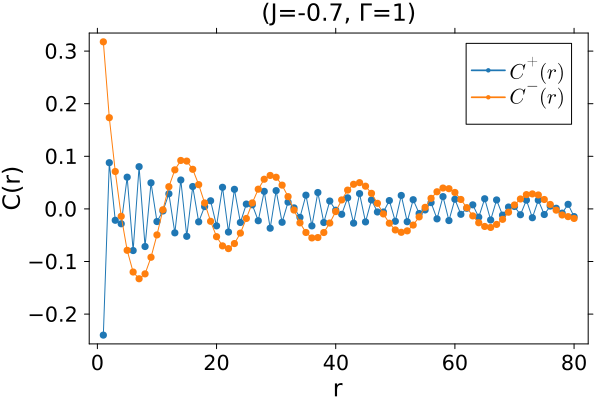}
 \captionsetup{justification=raggedright}
\caption{
Spin correlation function $C^+(r)$ and $C^-(r)$ as a function of site $r$ at a representative point $(J=-0.7,\Gamma=1)$ in the Gapless-I phase. 
DMRG numerics are performed on systems of $N=100$ sites using open boundary conditions.
The maximum bond dimension and truncation error in the calculations are set to \(m=1500\) and \(\epsilon=10^{-12}\), respectively.}
\label{fig.222}
\end{figure}

Next we examine the two-point spin correlation functions within the gapless phases. 
It is sufficient to focus on the Gapless-I phase, since the equivalence relation in Eq. (\ref{er1}) ensures that the two gapless phases are related by spin rotation $R(\hat{x},\pi)$. 
Fig. \ref{fig.222} show the DMRG numerical results for $C^+(r)$ and $C^-(r)$ at a representative point $(J=-0.7,\Gamma=1)$,
where $C^\pm(r)$ is defined in Eq. (\ref{correlation}). 
it is clear that both $C^{-}(r)$ and the envelope of $C^{+}(r)$ are oscillating with a wave vector incommensurate with the underlying lattice. Such behaviors are consistent with the exact solutions in Sec. \ref{sbsbsec:gapless} for the gapless phase,
as the wave vector $k_F$ in Eq. (\ref{eq:kF}) is incommensurate for general values of $(J,\Gamma)$, except special points.

\section{Summary}
\label{sec:summary}

In summary, we have studied the ground state properties of the exactly solvable spin-1/2 Kitaev–XX–Gamma chain. 
Using the Jordan–Wigner transformation and DMRG simulations, six phases are revealed, including four gapped phases with ferromagnetic or antiferromagnetic orders, and two gapless phases hosting helical Majorana fermions at low energies. 
The gapped phases have two-fold ground-state degeneracy and magnetic long-range order, while the gapless phases yield central charge $c =1$. 
Phase transitions include deconfined quantum critical lines with dynamical critical exponent $z = 1$ and quadratic critical points with $z = 2$, highlighting the rich interplay of symmetry, magnetic order, and quantum criticality.

%%%%%%%%%%%%%%%%%%%%%%%%%%%%%%%%%%%%%%%%%%%
\begin{acknowledgments}

W.Y. is supported by the National Natural Science Foundation of
China (Grants No. 12474476) and the start funding at Nankai University. 
DMRG calculations in this work were performed using the software package ITensor in  Ref. \onlinecite{ITensor}.
Numerical data is available upon requesting the authors. 

\end{acknowledgments}

\appendix

\begin{widetext}
%%%%%%%%%%%%%%%%%%%%%%%%%%%%%%%%%%%%%%%%%%%%%%%%%%%%%%
\section{Majorana operators}
The Majorana zero mode appears at the zero energy points $E_l(k_F)=0$. We can define the Majorana operators as
\begin{equation}
\begin{aligned}
     \alpha_{lk}^{(1)}&=\frac{1}{\sqrt{2}}(\gamma_{lk}+\gamma_{l,-k}^\dagger),\\
      \alpha_{lk}^{(2)}&=\frac{1}{\sqrt{2}i}(\gamma_{lk}-\gamma_{l,-k}^\dagger).
\end{aligned}
\end{equation}
It is straightforward to verify that they satisfy the Majorana condition in momentum space,
\begin{equation}
(\alpha^{m}_{lk})^\dagger=\alpha^m_{l,-k},
\end{equation}
as well as the canonical anticommutation relation
\begin{equation}
    \{(\alpha_{lk}^{m})^\dagger,\alpha_{l'k'}^{m'}\}=\delta_{ll'}\delta_{mm'}\delta_{kk'}.
\end{equation}
The model Hamiltonian takes the forms as 
\begin{equation}
    H= \sum_{k\in K_e}\sum_{l=\pm} i\,E_{l}(k) \alpha^{(1)}_{lk}\alpha^{(2)}_{lk}
\end{equation}
Computing the commutator between the Hamiltonian and these Majorana operators yields
\begin{equation}
\begin{aligned}
    [H,\alpha_{lk}^{(1)}]&=-iE_l(k)\alpha^{(2)}_{lk} \\
     [H,\alpha_{lk}^{(2)}]&=+iE_l(k)\alpha^{(1)}_{lk}
\end{aligned}
\end{equation}
At the zero–energy point, the Majorana zero mode emerges, since the corresponding operators no longer evolve in time.
\section{Proof the existance of zero energy point in gapless region}

When $\Gamma \neq 0$, we first determine the gapless regions. 
To determine the condition for $E_-(k_F)$ to vanish, we square the equation $E_-(k_F)=0$ and rearrange the terms,  yielding
\begin{equation}
    \Gamma^2 \sin^2(k/2) + J(J+1)\cos^2(k/2) = 0.
    \label{dd}
\end{equation}
 Since $\cos^2(k/2), \sin^2(k/2) \geq 0$ and $\Gamma^2>0$, a necessary condition for a solution of Eq. (\ref{dd}) to exist is
\begin{equation}
    J(J+1) \leq 0,
\end{equation}
giving the condition $-1 \leq J \leq 0$.

Eq. (\ref{dd}) can be further simplified to
\begin{equation}
    \Gamma^2 + \big[J(J+1)-\Gamma^2\big]\cos^2(k/2) = 0 \, ,
\end{equation}
which gives
\begin{equation}
    0 < \cos^2(k/2) = \frac{\Gamma^2}{\Gamma^2 + |J(J+1)|} < 1 \, .
\end{equation}
Thus, in the region $-1 \leq J \leq 0$, the dispersion always exhibits zero points at wave vectors  $k_F$ and $2\pi - k_F$, where
\begin{equation}
    k_F = 2\cos^{-1}\!\left(\sqrt{\frac{\Gamma^2}{\Gamma^2 + |J(J+1)|}}\right) \, .
    \label{eq:kF}
\end{equation}
\section{Proof the energy gap in the $(J>0,\Gamma>0)$ region}
In the region $(J>0,\,\Gamma>0)$, the lower branch of the spectrum can be written as
\begin{equation}
    E_-(u) = \sqrt{\Gamma^2 + D u^2} - \frac{u}{2}, 
\end{equation}
in which $D$ is defined as
\bea
D &=& \left(J+\tfrac{1}{2}\right)^2 - \Gamma^2,
\eea
and $u$ is given  by
\bea
u&=&|\cos(k/2)|.
\eea
To proceed, we need to distinguish between two scenarios. 

\begin{enumerate}

\item If the minimum of $E_-(u)$ occurs at $u_0\in(0,1)$ within the interior of the interval between $0$ and $1$, then $u_0$ must be a stationary point satisfying $E'_-(u_0)=0$. This condition gives
\begin{equation}
    0 < u_0^2 = \frac{\Gamma^2}{\,4D^2 - D\,} < 1, 
\end{equation}
yielding
\begin{equation}
4D^2 - D > \Gamma^2.
\end{equation}
In this  case,  the energy gap $\Delta_g$ is given by
\begin{equation}
    \Delta_g = \frac{\Gamma}{2}\sqrt{\frac{4D-1}{D}}.
\end{equation}

\item If instead the minimum of $E_-(k)$ is located at the boundary values $u_0=0$ or $u_0=1$, then the energy gap $\Delta_g$ is given by $E_-(u=0)=\Gamma$ or $E_-(u=1)=J$.
\end{enumerate}

Collecting both possibilities, the energy gap can be summarized compactly as
\begin{equation}
    \Delta_g = \min\left\{ J,\, \Gamma,\, \frac{\Gamma}{2}\sqrt{\tfrac{4D-1}{D}} \right\}.
\end{equation}

\end{widetext}
%%%%%%%%%%%%%%%%%%%%%%%%%%%%%%%%%%%%%%%%%%%%%%%%%%%%%%
%%%%%%%%%%%%%%%%%%%%%%%%%%%%%%%%%%%%%%%%%%%%%%%%%%%%%%

%%%%%%%%%%%%%%%%%%%%%%%%%%%%%%%%%%%%%%%%%%%%%%%%%%%%%%
%%%%%%%%%%%%%%%%%%%%%%%%%%%%%%%%%%%%%%%%%%%%%%%%%%%%%%

\end{document}